\documentclass[useAMS,usenatbib]{mn2e}
\usepackage{graphicx,fleqn,times,color}
\usepackage{lscape}
\usepackage{times}

\title[Dark matter and bar slowdown]{Bar slowdown and the distribution
  of dark matter in barred galaxies} 

\author[E. Athanassoula 
]{E. Athanassoula\thanks{E-mail:lia@lam.fr} \\
Aix Marseille Universit\'e, CNRS, LAM (Laboratoire d'Astrophysique de
Marseille),
UMR 7326, 13388, Marseille, France\\
}

\begin{document}

\date{Accepted 2013 November 13. Received 2013 November 11; in
  original form 2013 October 10}

\pagerange{\pageref{firstpage}--\pageref{lastpage}} \pubyear{2013}

\maketitle

\label{firstpage}

\begin{abstract}
`Conspiracy' between the dark and the baryonic mater prohibits an
unambiguous decomposition of disc galaxy rotation curves into the
corresponding components. Several methods
have been proposed to counter this difficulty, but their
results are widely discrepant. In this paper, I revisit 
one of these methods, which relies on the relation between the halo
density and the decrease of the bar pattern speed. The latter is
routinely characterised by the ratio ${\cal R}$ of the corotation
radius $R_{CR}$ to the bar length $L_b$, ${\cal R}=R_{CR}/L_b$. I use a
set of $N$-body+SPH simulations, including sub-grid physics, 
whose initial conditions cover a range of gas fractions and halo
shapes. The models, by 
construction, have roughly the same azimuthally averaged circular
velocity curve and halo density and they are all submaximal, i.e. according to
previous works they are expected to have all roughly the same ${\cal
  R}$  value, well outside the fast bar range (1.2 $\pm$ 0.2). Contrary to
these expectations, however, these simulations end up having widely different
${\cal R}$ values, either within the fast bar range, or well 
outside it. This shows that the ${\cal R}$
value can not constrain the halo density, nor determine whether galactic
discs are maximal or submaximal. I argue that this is true even for early type
discs (S0s and Sas).

\end{abstract}



\section{Introduction}
\label{sec:intro}

\subsection{Disc-to-halo mass degeneracy in disc galaxies}
\label{subsec:degeneracy}

Although extended HI rotation curves have clearly shown that there is
dark matter in disc galaxies \citep{Bosmathesis}, it is is still
unclear how massive this is and how it is distributed
\citep{Bosma.04}. Indeed, a rotation curve sets
constraints only on the total circular velocity,
but not on the contribution of each component individually at a given
radius. There is thus a degeneracy between the baryonic and the dark
mass, due to the fact that the mass-to-light ratio of the stellar
component ($M/L$) is poorly known. There is an upper limit to
the disc mass, since the contribution of the baryons can not exceed
the total circular velocity at any
radius, which is referred to as maximum disc. But 
the baryonic contribution could also be considerably less, in which
case the disc is referred to as submaximum.

Several methods have been proposed to solve this degeneracy.
Most of them rely on dynamical arguments, such as fitting the velocity
field of barred or spiral galaxies \citep[e.g.][]{LindbladLA96, KranzSR01,
  KranzSR03, WeinerSW01, PerezFF04, Zanmar.SWW08},
measurements of the stellar velocity dispersion \citep{Bottema93,
  KregelvdKF05, 
  Herrmann.Ciardullo.09, Bershady.MVWAS11, Martinsson.VWBAS.13}, the 
multiplicity of the spiral arms \citep{AthaBP87}, 
or the slow-down of the bar \citep[][hereafter DS98 and DS00,
  respectively]{Debattista.Sellwood.98, Debattista.Sellwood.00}. 
Non-dynamical arguments include the colour - $M/L$  
relation \citep{KassinJW06}, lensing \citep{Trott.W02, Tisserand.07,
  Hamadache.06, Trott.TKW10} and deviations from the 
Tully-Fisher relation \citep{Courteau.Rix.99, Gnedin.WPPR07}. Some of
these methods argue for maximum and others for submaximum discs, so
that no clear picture has yet emerged.

\subsection{Bar slowdown}
\label{subsec:slowdown}

Both analytical work \citep{Weinberg85} and simulations
(\citealt{Little.Carlberg.91}; \citealt{Hernquist.Weinberg.92}; DS98;
DS00; \citealt[][hereafter A03]{Athanassoula03};
\citealt{ONeill.Dubinski.03}; \citealt*{MartinezV.SH.06};
\citealt[][etc.]{Dubinski.BS.09}) have shown that the bar pattern
speed decreases during the evolution of a disc galaxy, sometimes very
considerably. This result, initially shown 
for purely stellar models, holds also for simulations with gas
\citep[][and references therein]{Berentzen.SMVH.07, Villa.VSH.10}. It
can be explained as due to the angular momentum exchange within the galaxy
or, alternatively but equivalently, by
the dynamical friction exerted by the halo on the bar,
which results in a decrease of the pattern speed. From both 
explanations, it is clear that a denser halo
will cause a more rapid bar slowdown, because it will
result in more halo mass in near-resonance locations, ready to absorb
angular momentum, and a stronger dynamical friction.
This link between bar slowdown and dark
matter was used by DS98 and DS00 to set constraints on the halo
mass and density and thus break the disc/halo degeneracy, in favour of
maximum discs.

It is not easy to compare pattern speed values of different
galaxies between them, or to
simulations, because the various galaxies and models have
different maximum rotational velocities and different total masses and
mass distributions. Instead, it is
customary to compare the corotation radius ($R_{CR}$), which
depends on the pattern speed, to the bar length ($L_b$). The ratio of these
two lengths, ${\cal R}=R_{CR}/L_b$, is thus often used as a yardstick for
measuring the pattern speed and it allows direct comparisons
between real galaxies and simulations. If $1 < {\cal R} < 1.4$ (or
1.5 for some authors) the bar is called fast, while if ${\cal R} > 1.4$
(1.5) it is considered slow.

A further difficulty comes from the fact that the pattern speed 
of bars in real galaxies can not be measured directly. The most widely
used  indirect method \citep{Tremaine.Weinberg.84} uses the luminosity
and the kinematics of the old stellar population along a slit and
provides a measure of the pattern relying  on the assumption that the
galaxy is in steady-state and the disc infinitesimally thin. 
This method was applied to a number of galaxies, as reviewed by 
\cite{Corsini.11}. Alternatively, the existence of
specific morphological features can give strong clues
to the location of the corotation radius 
\citep[e.g.][]{Rautiainen.SL.05, Rautiainen.SL.08, AthanassoulaRGBM10}.
\cite*{Perez.AMA.12} used outer rings, and thus were able
to study a much larger sample of galaxies than what had been
possible kinematically.

Dynamical arguments have also been used. Orbital structure theory
\citep{Contopoulos80} and resonant responses \citep{Atha80} show that
the bar can not extend beyond corotation, or, equivalently, that $\cal R
\ge$ 1. \cite{Atha92} used hydrodynamic simulations to show that, in
order for the shock loci in the bar to have the same shape as the
observed dust lanes, the pattern speed must be such as to fulfil
$R_{CR}=(1.2 \pm 0.2)L_b$ . This is the tightest constraint to date, but
it has not yet been tested to what extent the
tightness of these limits is model dependent. 
Note also that the above methods give information on the value of the
pattern speed at the present time, not on its decrease with time,
which would have been the most useful quantity for our
purposes. 

DS98 and DS00 used $N$-body simulations to set constraints on the halo
mass and found that, for their simulations, models
with a maximum disc have $1 \le {\cal R} \le 1.5$, while galaxies with
a submaximum disc (and therefore a halo of high density in the
central parts) have ${\cal R} > 1.5$. Taking into account the above described
dynamical and observational constraints on the ${\cal R}$ ratio, they
concluded that discs must be maximum.  

A03 first suggested that the value of $\cal{R}$ can not set
constraints on the halo-to-disc mass ratio in disc galaxies, because
it depends on the angular momentum redistribution in general, and
therefore on many quantities other than the disc-to-halo mass ratio.

In this paper I will re-examine the bar slow-down method and check whether it
can be used to set constraints to the amount and distribution of
dark matter. Sect. \ref{sec:simulations} presents the simulations
and Sect.~\ref{sec:results} the results. Discussion and conclusions
are given in Sect.~\ref{sec:discussion} and \ref{sec:Conclusions}, respectively.

\begin{figure*}
\includegraphics[scale=0.65,angle=-90]{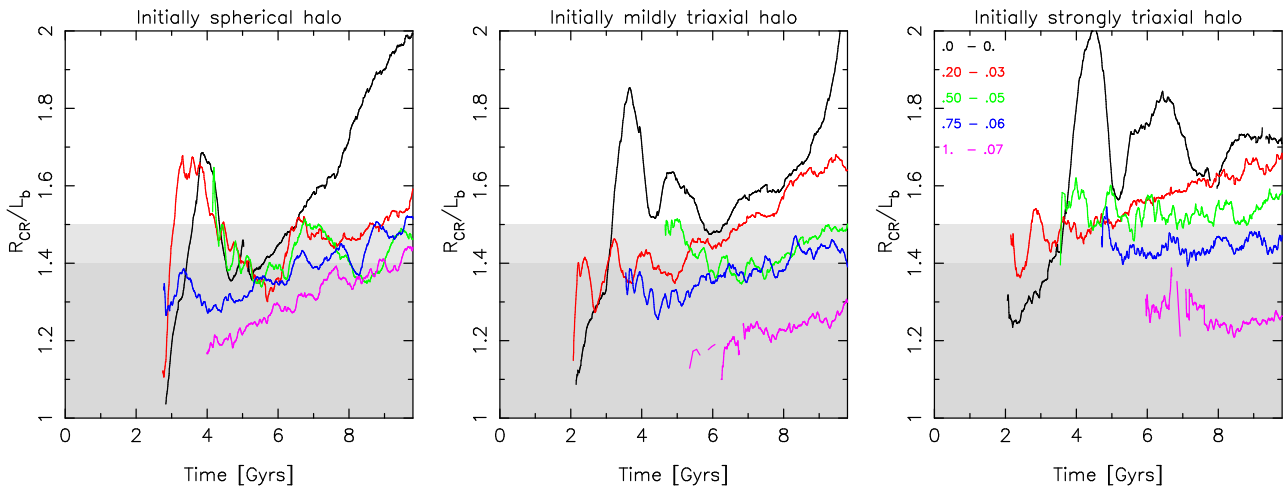}
\caption{The ratio $\cal R$=$R_{CR}/L_{b}$ as a function of time.
Simulations with initially spherical, mildly
triaxial and strongly triaxial haloes are given in the left, centre and
right panels, respectively. Different colours (in the on-line version)
denote 
simulations with different initial and final gas fractions, as noted in
the upper left corner of the right panel. The first of these two fractions is
the initial one (i.e. at $t$ = 0) and the second one is the final one
(i.e. at $t$ = 10 Gyrs). The darker (lighter) grey areas set the limit 
$1 \le~\cal{R}~\le$ 1.4 ($1.4 \le~\cal{R}~\le$ 1.5). 
} 
\label{fig:ratio}
\end{figure*}

\section{Simulations}
\label{sec:simulations}

I will here discuss fifteen simulations, including cases with
nonspherical haloes and/or with gas. They are described at length by
\citeauthor*{Athanassoula.MR.13} (\citeyear[][hereafter
  AMR13]{Athanassoula.MR.13}), where the reader can 
find all necessary information on the initial conditions and the run
parameters.  

Three different initial halo shapes are considered:
spherical, mildly triaxial (with axial ratios $b/a$ = 0.8 and $c/a$ =
0.6) and strongly triaxial ($b/a$ = 0.6 and $c/a$ = 0.4). 
These axial ratios increase during the evolution 
and end up much more spherical. 
Furthermore, five different values of the initial gas fraction in the
disc (0, 20, 50, 75 and 100\%) are considered, but, 
since star formation is included in the simulations,
these values change with time, reaching after 6 Gyrs values between 4
and 9\%, and after 10 Gyrs values between 3 and 7\% of the disc mass
(Fig. 2 of AMR13).  

Initially, all simulations have roughly the same azimuthally averaged halo
and total circular velocity curves. The maximum 
contribution of the disc to the total circular velocity curve is
roughly equal to that of the halo at the same radius, i.e. the disc
is clearly submaximum with a mean value of the Sackett parameter
\citep{Sackett.97} roughly equal to 0.7. More precisely,
averaging over all runs I get $V_d/V_{t}$ = 0.68 $\pm$ 0.01,  
where $V_d$ and $V_t$ are the circular velocity of the disc and the
total circular velocity, both
measured at the location where the disc's contribution
is maximum. From figure 2 of DS98, one would then expect that all
these simulations will have $\cal R > $ 2.5.

\section{Results}
\label{sec:results}

Fig.~\ref{fig:ratio} shows the ratio $\cal R$ as a function of time
for the fifteen simulations. The curves do not start out from $t$ = 0
because the bar must have grown considerably before the
pattern speed can be safely measured.
In a number of the curves, particularly those
corresponding to simulations which have both a high initial gas
fraction and initial triaxiality, there are clear gaps. These
correspond to time ranges during which the bar
length is particularly difficult to measure, so I preferred to leave
out the corresponding estimates (see Appendix).

\begin{figure}
\includegraphics[scale=0.7,angle=-90]{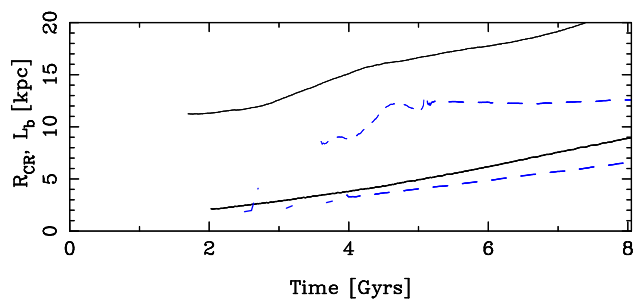}
\caption{Corotation radius (black solid line) and bar length (blue dashed line)
  as a function of time for two runs with identical spherical haloes. The upper
  two lines correspond to a run with no gas and the two lower ones to
  a run with initially 100\%gas. Note the difference in lengths
  between the two runs, but also the difference in the relative
  increase of the corotation radius with respect to the bar length.
} 
\label{fig:R(t)}
\end{figure}

The general evolutionary trend is an increase of $\cal R$ with
time. This increase varies from very strong (as in the gas-less
cases), to 
practically zero. The latter is seen in cases with a strong initial
triaxiality and an initial gas fraction between 50 and 100\%, i.e.
between 5 and 7\% at $t$=10 Gyrs.
This behaviour is explained in Fig.~\ref{fig:R(t)}, where 
I plot both the (smoothed) corotation radius and bar length as a function of
time for two runs, one gasless and the other with initially 100\% gas,
but having the same haloes. As expected, the gasless simulation has too
large corotation and bar lengths, while the initially very gas rich one
has very realistic values. This, together with its implications, will
be discussed at length elsewhere. This plot also explains the relatively 
low values of $\cal{R}$ for gas rich versus gas poor cases. It thus becomes
clear that in models
with a low gas fraction and no or low triaxiality the  
corotation radius increases much more rapidly than the bar length, so
that their ratio $\cal R$ increases, sometimes very considerably. In
contrast, for models with high gas fraction and triaxiality, the
increase of $R_{CR}$ is not much stronger than that of the bar length,
so that the $\cal{R}$ ratio increases little, if at all, with
time. No models had a corotation radius increasing less than the bar
length over a long period of time, i.e. in no models did the $\cal{R}$
ratio decrease globally over the evolution.

\begin{figure}
\includegraphics[scale=0.4,angle=-90.]{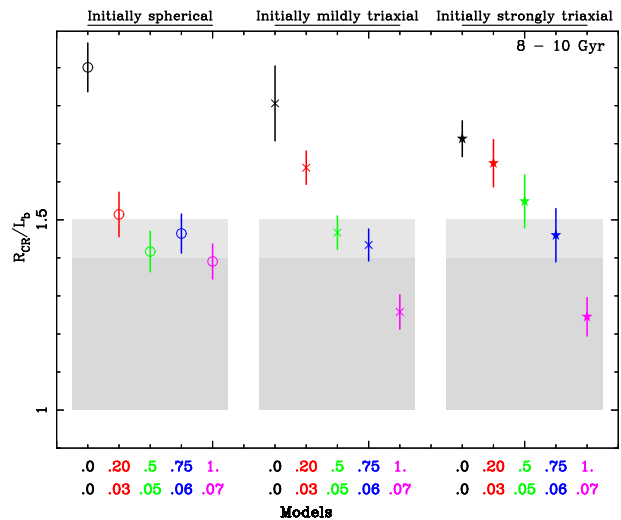}
\caption{Average and dispersion of $\cal R$ in the time range 8 to 10
  Gyrs, colour-coded as in Fig.~\ref{fig:ratio}. The dispersions
  provide an indication of the amount of evolution occurring 
  in the adopted range of time. The values in the
  upper (lower) line of the $x$-axis label give the gas fraction at
  $t$=0 ($t$=10) Gyrs. The simulations are
  divided in three blocks according to their halo shape. Simulations
  with initially
  spherical (left), mildly triaxial (centre) and strongly triaxial
  (right) haloes are plotted with open circles, X symbols and
  filled stars, respectively.
} 
\label{fig:fast_slow_global}
\end{figure}

A global view of these results is given in
Fig.~\ref{fig:fast_slow_global}. We plot the averages of $\cal{R}$ in
a time range towards the end of the simulation, namely between 8
and 10 Gyrs, in a way that displays
clearly the two main results of this study. 
First, the $\cal{R}$ values depend very strongly on the gas
fraction in the simulation. For simulations with
no gas they are quite high, and they decrease as we consider simulations
with a larger gas fraction. This is clear 
for all three halo triaxialities. All simulations with initially
100\% or 75\% gas (7\% or 6\% at $t$=10 Gyrs), as well as the majority of those with
50\% gas (5\% at $t$=10 Gyrs), are in the range
(1., 1.5). But most of the remaining cases, and in  particular the
cases with no gas, are well outside  
these limits. Note also that the minimum values of $\cal R$ are around 1.2, and that no
case with $\cal{R} <$ 1 has been found. 

The above lead to the second result, which is the most important in
the context of this paper, namely that, although initially the
radial profile of the disc-to-halo mass ratio is roughly the same for  
all models shown here, their $\cal{R}$ ratios differ very
considerably, some corresponding to fast and others to slow bars. 

\section{Discussion}
\label{sec:discussion}

There are a number of differences between the simulations of DS00 and
those of AMR13, which can explain the differences between our
conclusions. A fair number of these differences are due to the
relatively restricted computer power available 10 - 15 years ago. In
order of increasing importance these are:  
First, the simulations of DS00 have a much
lower resolution (larger softening) than those discussed here.   
Second, the initial conditions of the AMR13 simulations were
built specifically so as to be near axisymmetric equilibrium (i.e. in
good equilibrium for timescales up to bar formation times). In 
contrast, the simulations in DS98 and DS00 generally start off with
Toomre's parameter $Q$ = 0.05, i.e. they are strongly axisymmetrically unstable
which will drive a strong reorganisation of the disc 
material with corresponding transients and heating. Third, the DS98 and DS00
submaximum disc rotation curves
drop too steeply after the maximum to be realistic
(e.g. Fig. 1 of DS98 and 2 and 8 
of DS00), in contrast to ours, shown in Fig. 1 of AMR13. Fourth and
most important, our simulations consider a larger variety of models, including
gas and its physics, as well as nonspherical haloes. 

Both halo nonaxisymmetry and particularly the presence of gas occur naturally in
disc galaxies and thus need to be included in numerical simulations
before any conclusions on constraints on halo mass can be reached.  
Cosmological simulations show that haloes form triaxial,
with axial ratios compatible with those assumed in our initial
conditions \citep[e.g.][and references
  therein]{Vera-Ciro.SHFNSVW.11}. Then the formation of discs and bars
decreases this 
triaxiality and brings it to levels compatible with the present
observational constraints. The presence of gas in galactic discs is
also to be expected. It is dominant in 
these discs early on and its fraction decreases with time due to star
formation, to reach present day levels at redshift $z$ = 0. I showed
here that both the halo nonaxisymmetry and particularly the presence
of gas lead to a reduction of $\cal R$ values, bringing some slow bar
cases within the fast bar range. 

All theoretical works presented so far -- and independent of the method
used (orbital structure
theory, response calculations, or simulations) -- agree well on the fact
that $\cal{R}$ should be larger than unity, i.e. that the whole bar
should lie within corotation. Yet observations have come up with a
number of galaxies where this is not the case \citep{Corsini.11,
  Perez.AMA.12}. Either some substantial 
physics is missing from all the theoretical approaches, or, 
more likely, the error bars of the observations are
underestimated. By this I mean not only the statistical errors,
e.g. due to a wrong estimate of the viewing angles,
but also biases, such as can be due to time evolution or to the 
existence of out of the plane motions in the bar.
  
The simulations used in this study have different gas
fractions and different halo shapes, while the remaining quantities 
were kept constant across
the models. It is thus not possible to draw from them conclusions about
the variations of the $\cal R$ ratio across the Hubble
sequence. Indeed, late type disc galaxies are not only more gas rich than early
types, but they are also less massive, have a relatively less massive
classical bulge and a less steeply rising rotation curve. Thus one can not,
based only on 
the results presented here, draw conclusions on how $\cal{R}$ varies
as a function of galaxy type, nor confirm or invalidate the
results found by \cite{Rautiainen.SL.05,Rautiainen.SL.08}.
 
A ratio $\cal R$ which is near-constant
with time, does not does not necessarily imply that there is no
evolution, and no increase of the
corotation radius or bar length with time. Indeed, it is possible that
both $R_{CR}$ and $L_b$  
increase with time in such a way that their ratio stays roughly constant. 
For example, in runs with an initial gas fraction between 50 and 100\%
and a strongly triaxial halo, 
${\cal R}$ stays roughly constant with time, yet both the bar length and
corotation radius increase steadily with time. 
Similar behaviour has been found in a number of other
simulations, e.g. as displayed in Fig. 4 of \cite{MartinezV.SH.06} for
a gasless case.

The results found here concern also S0 and Sa galaxies, despite the fact that
these galaxies have little gas, because their discs, like those of all
other disc galaxies, must
have formed from gas, so must have been at earlier times quite
gas rich. Furthermore, it is not clear at what stage of the bar
formation and evolution process their gas was expelled or turned into
stars, so that the bar evolution could well have been influenced by
it. For example the simulations in AMR13 lost most of their gas and of
their triaxiality early on during their evolution and yet their
gas influences the ${\cal R}$ ratio sufficiently to make the big
difference we can see in Figs.~\ref{fig:ratio} and \ref{fig:fast_slow_global}. 

Even more important,
although I focused here on the gas content and halo shape, 
these two properties should not be considered as the only two (other
than the halo mass distribution) to influence the bar slowdown. Any
property influencing 
the redistribution of angular momentum within the galaxy will also
influence the bar slow down. As already discussed in A03, this
includes a number of other properties such as the velocity dispersion
of the stellar disc, the possible existence of a classical bulge and its mass
distribution, whether the spheroidal components rotate, and
if yes how much, the velocity dispersion in these spheroidals, etc. 
It is thus not possible to use the value of $\cal{R}$ to set constraints
on the halo density in the inner parts, even 
for S0s. This is particularly important since most
of the observational measures of $\cal R$ using kinematics 
have been made on early type, high surface brightness barred galaxies.

\section{Conclusions}
\label{sec:Conclusions}

In this paper I examined one of the methods often used to argue against
submaximum discs (and thus indirectly against cuspy halo profiles as
in \citealt{Navarro.FW.96}), using a more general class of simulations
than those often used so far, namely simulations including gas and its
physics and halo triaxiality. I find a clear trend between the
value of the $\cal{R}$ ratio and the fraction of the disc mass which
is in gas. Namely, for smaller amounts of gas I find that both the bar
length and the corotation radius increase stronger with time, in such
a way that globally  
$\cal R$ also increases. Most important, I find that for roughly the
{\it same} halo-to-disc mass ratio profiles, the $\cal{R}$ value can,
depending on the gas fraction and the halo shape, be within either the
fast or the slow bar region. 
In particular, I find several submaximum disc models which have an
$\cal{R}$ value which was so far thought to correspond only to maximum
disc models. Thus the value of $\cal{R}$ can not be used to discriminate between 
maximum and submaximum discs. Hence one of the arguments very often used in
favour of near-maximum discs is dismissed. This does not mean that
discs are submaximum, it just means that this method can not solve
the problem and that more work is necessary to
break the disc/halo degeneracy.


\section*{Acknowledgements}

I thank A. Bosma and S. Rodionov for useful and
stimulating discussions, the referee for helpful comments and J.-C. Lambert for computer
assistance. I acknowledge financial support to the DAGAL network 
from the European Union's Programme FP7/2007-2013/ under REA grant 
agreement number PITN-GA-2011-289313.

\bibliography{athanassoula_omp.bbl}

\appendix
\section[]{Measuring the corotation radius and the bar length}
\label{sec:rcorot-barlength}

For all the runs I calculated the pattern speed as a function of time
and from this and the mass distribution at the corresponding time, I
obtained $R_{CR}$ for all times after the bar has grown
sufficiently for its angle to be accurately measured,. 

\begin{figure}
\includegraphics[scale=0.7,angle=-90]{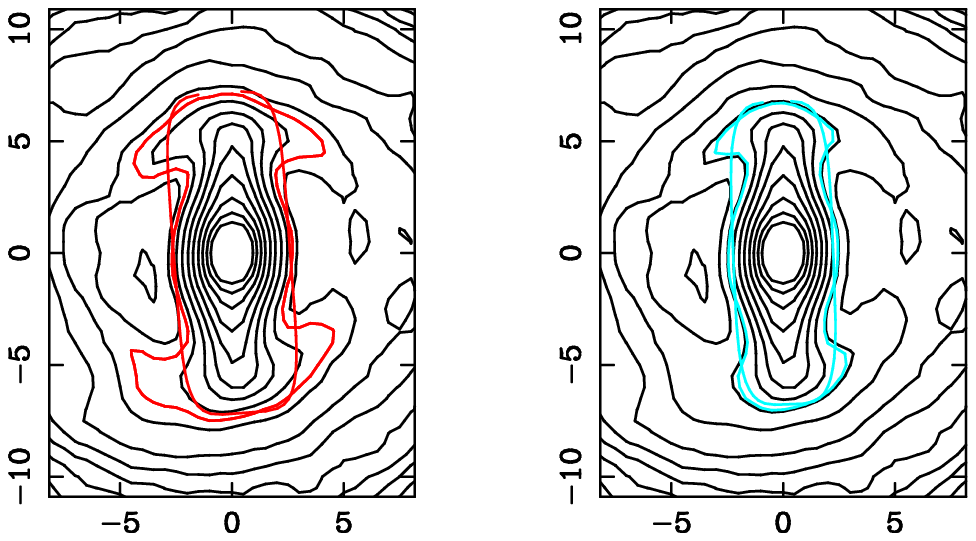}
\caption{Measuring the bar length in a snapshot, using the results of
  generalised ellipse fits (left panel), or the angle of the $m$ = 2
  Fourier component (right panel). The isophote corresponding to the
  end of the bar, as well as the corresponding best fitted generalised
  ellipse is also given in red (blue) for the two methods. 
} 
\label{fig:barlength}
\end{figure}

The workhorse for our measurements of the bar length is a method very
similar to what was used in DS00 and in
\citeauthor{Athanassoula.Misiriotis.02} (\citeyear[][hereafter 
  AM02]{Athanassoula.Misiriotis.02}), i.e. the surface of the disc 
was cut in annuli and in each one of these I calculated the phase of
the $m$=2 component. The radius at which this angle deviated
significantly from that of the bar is a measure of the bar length (DS00,
AM02 etc.). I also used two further methods described in AM02, which
rely on the fit of generalised ellipses \citep{Athanassoula.MWPPLB.90} to the isophotes.
This fit gives us also the angle of the isophote, from
which the bar length can be deduced as the radius within which this angle
does not vary significantly. Third, if the bar is sufficiently
strong, the ellipticity as a function of radius shows a clear drop,
which can be used as a measure of the bar
length (see e.g. the upper left panel of figure 4 of
AM02), as already discussed in AM02 and \citealt{Gadotti.ACBSR07}.
In the latter work, this third method was compared to other methods
using real galaxies  
and was found to give quite satisfactory results. Fourth, we included
eye estimates of the bar length for a number of randomly chosen
snapshots -- as a test of 
the various methods -- as well as for the times when one or more of
the methods presented a problem. The finally 
adopted bar length is the average of all results which did not present
any problem. For snapshots which presented problems for several
methods I did not adopt any result: since a snapshot was saved 2000
times per run, some of these times could be easily neglected. 

The results of the first and third method, which rely on very
different properties of the bar, are compared for a typical snapshot in
Fig.~\ref{fig:barlength}. It shows that a different isophote
was picked out by each of the two methods, the ellipticity drop method
leading systematically to somewhat longer bars. 
This difference is, however, relatively small due to the fact 
that the drop of the density at the bar end along the direction of the
bar major axis is quite abrupt, so that exactly which isodensity is
picked does not change the results substantially. 
Note that if I use ellipses 
rather than generalised ellipses, this difference becomes much more
important because the ellipse shape can not describe
adequately the end of the bar, contrary to the generalised ellipse,
where this is easily achieved. This inadequacy has to be kept in mind
when applying this 
method to large observational samples. A complete discussion on the
calculations of bar lengths will be given elsewhere.

\label{lastpage}

\end{document}